\documentstyle[12pt,graphicx]{article}

\title{Magnus Force and Aharonov-Bohm Effect 
in Superfluids}
\author{Edouard Sonin \\
Racah Institute of Physics,\\ Hebrew University of
Jerusalem \\and \\Ioffe Physical Technical Institute, Solid State Physics 
Division,\\ St. Petersburg, Russia}
\date{\today}
\begin{document}

\maketitle              

\begin{abstract}
The paper addresses the problem of the transverse force (Magnus 
force) on a vortex
in a Galilean invariant quantum Bose liquid. Interaction of 
quasiparticles
(phonons) with a vortex produces an additional transverse force 
(Iordanskii
force). The Iordanskii force is related to the acoustic Aharonov--Bohm 
effect.
Connection of the effective Magnus force with the Berry phase is also 
discussed.
\end{abstract}

\section{Introduction} \label{In}

In classical hydrodynamics it has long been known that  if the vortex 
moves 
with  respect to a liquid there is a force on the vortex normal to the
vortex velocity  \cite{Magnus}. This is the Magnus force, which
is a particular case of a force on a body immersed into a liquid with 
a
flow circulation around it (the Kutta-Joukowski theorem). An
example of it is  the lift force on an airplane wing.

The key role of the Magnus force in vortex dynamics became clear from 
the
very beginning of studying superfluid hydrodynamics. In the pioneer 
article
on the subject Hall and Vinen \cite{HV} defined the
superfluid Magnus force as   a force between a vortex and a
superfluid. Therefore it was proportional to the superfluid density 
$\rho_s$. 
But in the two-fluid hydrodynamics the superfluid Magnus force is not 
the only
force on the vortex transverse to its velocity: there was also a 
transverse
force between the vortex and quasiparticles moving with respect to 
the vortex.
The transverse force from rotons was found by Lifshitz and Pitaevskii
\cite{LP} from the quasiclassical scattering theory. Later  Iordanskii
\cite{I} revealed the transverse force from phonons. The
analysis done in Ref. \cite{S} (see also Refs. \cite{RMP,PRB7}) 
demonstrated
that  the Lifshitz--Pitaevskii force for rotons and the Iordanskii 
force for
phonons originate from interference between quasiparticles which move 
past the
vortex on the left and on the right sides with different phase 
shifts, like
in the Aharonov--Bohm effect \cite{AB}. Since the phase shift depends 
on the
circulation which is a topological charge for a vortex, this is a 
clear
indication of connection between the transverse quasiparticle force 
and
topology. 

Later on the analogy between wave scattering by vortex and electron 
scattering by
the magnetic-flux tube (the Aharonov--Bohm effect)  was studied  in 
classical 
hydrodynamics for water surface waves (the acoustic Aharonov--Bohm 
effect 
\cite{Berry,R}). Scattering of  the light by a vortex also results in 
the
Aharonov--Bohm interference (the optical Aharonov--Bohm effect 
\cite{LePi}). As
follows from Ref. \cite{S}, the Aharonov--Bohm interference always 
produces a
transverse force on the vortex, or the fluxon. For the original 
Aharonov--Bohm
effect this force is discussed by Shelankov \cite{ShelB}. 

The Magnus force on a vortex in a superconductor was introduced by 
Nozi\`eres
and Vinen \cite{NV}.  The total transverse force on a vortex is 
responsible for
the Hall effect in the mixed state. In superconductors not
only quasiparticles, but also impurities produce an additional 
transverse
force on the vortex. A reader can find discussion of this problem in 
the
review by Kopnin \cite{KK-S}.  

Despite a lot of work done to understand and calculate the Magnus 
force, it
remained to be a controversial issue. Ao and Thouless  \cite{AT} 
pointed out a connection of the Magnus force with the Berry phase
\cite{B} which is the phase variation of the quantum-mechanical wave
function resulting from transport of the vortex round
a close loop. From the Berry-phase analysis  Ao and Thouless 
concluded that
the  effective Magnus force is proportional to the superfluid density,
and there is no transverse force on the vortex from quasiparticles and
impurities \cite{AT,AT-N,GWT}. Such conclusion disagreed with the 
previous
calculations of the Magnus force in superfluids and
superconductors, and therefore generated a vivid discussion
\cite{Disc,ComS,comR}.

The present paper addresses this problem \cite{SCam} which is very
important for many issues in modern condensed matter physics, field 
theory,
and cosmology. The analysis is based on studying momentum balance in 
the area
around a moving vortex without using any preliminary concept of a
force. The word {\em force} is a label to describe a transfer of 
momentum between
two objects. A careful approach is to give these labels to the 
various terms in
the momentum balance equation only {\em after} derivation of this 
equation. 
I restrict myself with the problem of the Galilean invariant quantum 
Bose
liquid. For a weakly nonideal Bose gas one can use
the Gross--Pitaevskii theory \cite{GP}. In this theory the liquid
is described by the nonlinear Schr\"odinger equation.  At large
scales the nonlinear Schr\"odinger equation yields the
hydrodynamics of an ideal inviscous liquid. In presence of  an 
ensemble of sound
waves (phonons)  with the Planck distribution, which is characterized 
by a locally
defined normal velocity $\vec v_n$ (the drift velocity of the Planck 
distribution), 
one obtains two-fluid hydrodynamics. Eventually the problem of the 
vortex motion in
presence of the phonon normal fluid is a  problem of hydrodynamics.

The paper starts from discussion of the Magnus force in classical
hydrodynamics (Sec. \ref{cl}). In Sec. \ref{supMag} I define
the superfluid Magnus force and the force from quasiparticles 
scattered by a
vortex. Connection between the Gross--Pitaevskii theory and the 
two-fluid
hydrodynamics is discussed in Sec. \ref{Schrod}. In Sec.
\ref{Born} I discuss scattering of a sound wave  by a vortex
and show that the standard scattering theory fails to give a 
conclusive
result on the transverse force because of small-scattering-angle 
divergence
of the scattering amplitude. The next
Sec. \ref{AB-I} gives a solution of the sound equation around the 
vortex,
which is free from the small-angle divergence. Using this solution in 
the
momentum balance one obtains the equation of vortex motion, which 
contains
the Iordanskii force. The momentum transfer responsible for the 
Iordanskii
force occurs at small scattering angles where a phenomenon analogous 
to the
Aharonov--Bohm effect is important: an interference between the waves 
on the
left and on the right from the vortex with different phase shifts 
after
interaction with the vortex. Section \ref{AB-PW} shows how to derive 
the
transverse force from the exact partial-wave solution of the 
Aharonov--Bohm
problem for electrons. In our analysis the force on the vortex 
originates from
scattering of noninteracting quasiparticles. This is a valid 
assumption since
normally the phonon mean-free path essentially exceeds the scale 
where the
force arises (the phonon wavelength) \cite{Cr-T}. But in order to 
know the
effect of this force on the whole superfluid, it is important to 
investigate
how the force is transmitted to distances much larger than the 
mean-free path
where the two-fluid hydrodynamics becomes valid. It is done in Sec. 
\ref{2D}.
The effect of the force at very large distances is also important for
discussion of the Berry phase in Sec.\ref{Berr}. The transverse force 
creates
a circulation of the normal velocity at very large distances from an 
isolated
vortex. Taking into account this circulation, the Berry phase yields 
a correct
value of the transverse force, which agrees with the result  derived 
from the
momentum balance.

The present paper does not discuss experimental aspects of this 
problem, which
are addressed in other reviews devoted to rotating $^4$He and $^3$He
\cite{RMP,JK,Bevan}).

\section{The Magnus force in classical hydrodynamics} \label{cl}

It is worth to remind how the Magnus force appears in classical
hydrodynamics. Let us consider a cylinder immersed  in an
incompressible inviscous liquid. There is a potential circular flow 
around the
cylinder with the velocity 
\begin{equation}
\vec v_v(\vec r)=\frac{\vec \kappa \times \vec r}{2\pi r^2}~.
   \label{v-v} \end{equation}
Here $\vec r$ is the position vector in the plane $xy$, the axis $z$ 
is the
axis of the cylinder, and $\vec
\kappa$ is the circulation vector along the axis $z$. In classical
hydrodynamics the velocity circulation $\kappa = \oint \vec v_v \cdot 
d\vec
l$ may have arbitrary values. There is also a
fluid current past the cylinder with a transport velocity $\vec 
v_{tr}$ and the
net  velocity is 
\begin{equation}
\vec v(\vec r) = \vec v_v(\vec r) + \vec v_{\rm{tr}}~.
         \label{vel} \end{equation}
This expression is valid at distances $r$ much larger than the 
cylinder
radius. At smaller  distances one should take into account that the 
flow with
the velocity $\vec v_{tr}$ cannot penetrate into the cylinder, but 
velocity
corrections due to this effect decrease as $1/r^2$ and are not 
essential for
the further analysis.

The Euler equation for the liquid is
\begin{equation}
\frac{\partial \vec v}{\partial t} + (\vec v \cdot \vec \nabla)\vec 
v= - {1
\over \rho} \vec \nabla P ~.
   \label{Euler} \end{equation}
Here $\rho$ is the liquid density and $P$ is the pressure. 

\begin{figure}
\begin{center}
\includegraphics[width=.7\textwidth]{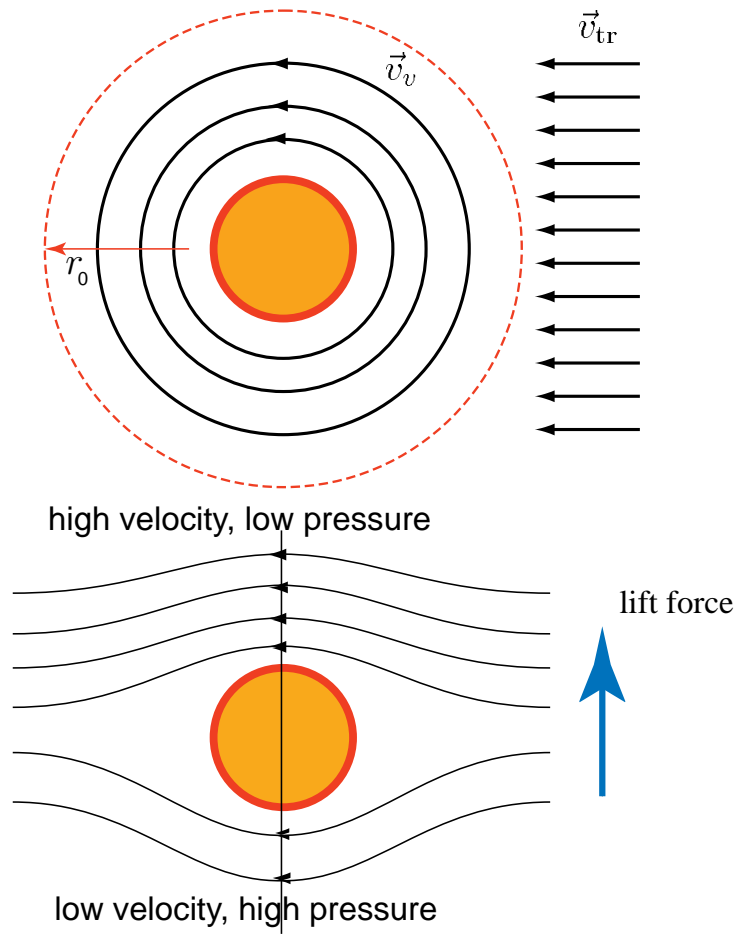}
\end{center}
\caption[]{Magnus (lift) force. It is derived from the momentum 
balance inside
a cylinder of radius $r_0$.}
\label{f1}
\end{figure}

Assuming that the cylinder moves with the constant velocity $\vec 
v_L$,
i.e., replacing the position vector $\vec r$ by $\vec r - \vec v_L 
t$, one
obtains that 
\begin{equation}
\frac{\partial \vec v}{\partial t}= - (\vec v_L \cdot \vec 
\nabla)\vec v~.
       \label{time-der} \end{equation}
Then the Euler equation (\ref{Euler}) yields the Bernoulli law for the
pressure: 
\begin{equation}
P= P_0 -{1 \over 2} \rho [\vec v(\vec r) - \vec v_L]^2 =P_0'
-{1 \over 2} \rho \vec v_v(\vec r)^2 -\rho \vec v_v(\vec r) \cdot 
(\vec v_{\rm{tr}} -
\vec v_L)~.
              \label{Bernu} \end{equation}
Here $P_0$ and $P_0'=P_0 -{1 \over 2} \rho (\vec v_{\rm{tr}} -\vec 
v_L)^2$
are constants, which are of no importance for the following 
derivation.
Figure \ref{f1} shows that due to superposition
of two fluid motions given by Eq. (\ref{vel}) the velocity above the
cylinder is higher than below the cylinder. According to  the 
Bernoulli law, the pressure is higher in the area where the velocity 
is
lower. As a result of it, a liquid produces a force on the cylinder 
normal
to the relative velocity of the liquid with respect to the cylinder. 
This
is a lift force, or the Magnus force.    

In order to find the whole force, we must  consider the
momentum balance for a cylindrical region of a radius $r_0$ around the
cylinder (see Fig. \ref{f1}). The momentum conservation law requires 
that the
external force $\vec F$ on the cylinder is equal to the momentum flux
through the entire cylindrical boundary in the reference frame moving 
with
the vortex velocity $\vec v_L$. The momentum-flux tensor is
\begin{equation}
\Pi_{ij} =P\delta_{ij} + \rho  v_i(\vec r) v_j(\vec r)~,
    \label{mom-flux} \end{equation}
or in the reference frame moving with the vortex velocity $\vec v_L$:
\begin{equation}
\Pi'_{ij} =P\delta_{ij} + \rho  ( v_i-  v_{Li})  (v_j -v_{Lj})~.
    \label{mom-flux-L} \end{equation}
The momentum flux through the  cylindrical surface of radius $r_0$ is 
given
by  the integral $\int dS_j \Pi'_{ij}$ where
$dS_j$ are the components of the vector $d\vec S$ directed along the 
outer
normal to the boundary of the cylindrical region and equal to the 
elementary
area of the boundary in magnitude. Then using Eqs. (\ref{v-v}),
(\ref{Bernu}), and (\ref{mom-flux-L}), the momentum balance yields the
following relation: 
\begin{equation}
\rho [(\vec v_L - \vec v_{\rm{tr}}) \times \vec \kappa] =\vec F~. 
   \label{Magnus} \end{equation}

On the left-hand side of this equation one can see the Magnus force 
as  it
comes in the classical hydrodynamics. The Magnus
force balances the resultant external force $\vec F$ applied to the 
cylinder.
In the absence of external forces the cylinder moves with the 
transport
velocity of the liquid: $\vec v_L=\vec v_{\rm{tr}}$ (Helmholtz's 
theorem).

In the derivation we used the hydrodynamic equations only at large
distance from the vortex line, and the radius of the cylinder does 
not appear
in the final result. Therefore Eq. (\ref{Magnus}) is valid even if the
circular flow with circulation $\kappa$ occurs without any cylinder 
at all.
In hydrodynamics such a flow pattern is called a vortex tube, a 
vortex line,
or simply a vortex. Hydrodynamics is invalid at small distance from 
the
vortex line. This area is called the vortex core.
But this does not invalidate the derivation of the Magnus force for a 
Galilean
invariant liquid, in which the momentum is a well-defined {\em 
conserved}
quantity even inside the vortex core where the hydrodynamic theory 
does not
hold.

\begin{figure}
\begin{center}
\includegraphics[width=.9\textwidth]{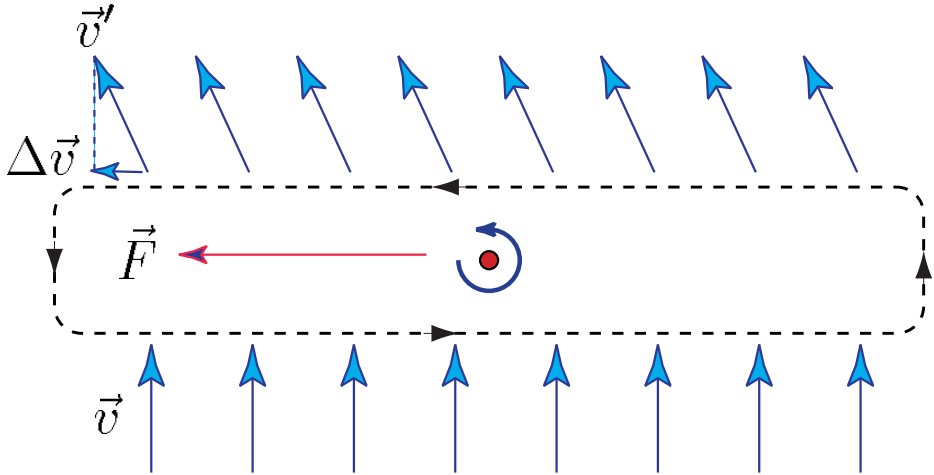}
\end{center}
\caption[]{Momentum balance in a stripe area. The transverse force is
determined by the current  circulation
$\rho \oint  \Delta \vec v \cdot d\vec l$ round the stripe.}
\label{f1a}
\end{figure}

In the momentum balance in a cylinder around a vortex, a half of the
Magnus force is due to the Bernoulli contribution to the pressure, 
Eq. (\ref{Bernu});
another half is due to the convection term
$\propto v_iv_j $ in the momentum flux. However, such decomposition 
of the resultant
momentum flux onto the Bernoulli and the convection parts is not 
universal and
depends on the shape of the area for which we consider the momentum 
balance. We may
consider the momentum balance in a stripe, which contain the vortex 
inside and is
oriented normally to the transport flow (see Fig. \ref{f1a}). This 
yields again Eq.
(\ref{Magnus}), but now the pressure (the Bernoulli term) does not 
contribute to the
transverse force at all, and only convection is responsible for the 
Magnus force.
Then the physical origin of the Magnus force is the following. The 
liquid enter the
stripe, which contains a vortex, with one value of the transverse 
velocity (equal to
zero in Fig. \ref{f1a}) and exits from the stripe with another value 
of it. A difference between two values is $\Delta
\vec v$ in Fig. \ref{f1a}. The transverse force is the total 
variation of the
transverse (with respect to the incident velocity $\vec v$) liquid 
momentum per unit
time. The latter is equal to a product of the current circulation
$\rho \oint  \Delta \vec v \cdot d\vec l$ and the velocity $v$.

\section{The Magnus force in a  superfluid}
\label{supMag}

In the superfluid state liquid motion is described by two-fluid
hydrodynamics: the liquid consists of the superfluid
and  the normal component with the superfluid and the normal density
$\rho_s$ and $\rho_n$, and the superfluid and the normal velocity 
$\vec
v_s$ and $\vec v_n$ respectively. The circular motion around the 
vortex
line is related to the superfluid motion. Therefore Hall and Vinen 
\cite{HV}
suggested that the Magnus force is  entirely connected with the
superfluid density $\rho_s$ and the superfluid velocity $\vec v_s(\vec
r)=\vec v_v(\vec r)+ \vec v_{s\rm{tr}}$. Then instead of Eq. 
(\ref{Magnus})
one has:       
\begin{equation} 
\rho_s [(\vec v_L - \vec v_{s}) \times \vec \kappa]
=\vec F~. 
   \label{Magnus-S} \end{equation}
Here and later on we omit the subscript ``tr'' replacing $\vec 
v_{s\rm{tr}}$
by $\vec v_s$. But one should remember that the superfluid velocity 
$\vec v_s$
in the expression for the Magnus force is the superfluid velocity far
from the vortex line.

The ``external'' force $\vec F$ is in fact external not for the whole 
liquid, but only for its superfluid part. The force appears due to 
interaction
with quasiparticles which constitute the normal  part of the liquid, 
and
therefore is  proportional to the relative velocity $\vec v_L - \vec 
v_n$
\cite{HV}. For a  Galilean invariant liquid  with  axial symmetry  
the most
general expression for the force
$\vec F$ is
\begin{equation}
\vec F = -D(\vec v_L - \vec v_n) - D'[\hat z \times (\vec v_L - \vec
v_n)]~.
     \label{F} \end{equation}
The force component $\propto D'$ transverse to the velocity $\vec v_L 
-
\vec v_n$  is possible because of broken time invariance in presence 
of a
vortex and resulting asymmetry of quasiparticle scattering by a 
vortex.

Inserting  the force $\vec F$ into Eq. (\ref{Magnus-S}), one can 
rewrite the 
equation of vortex motion collecting
together the terms proportional to the velocity $\vec v_L$:  
\begin{equation}
\rho_M [\vec v_L \times \vec \kappa] + D \vec v_L = \rho_s [\vec v_s 
\times \vec \kappa] 
  +D \vec v_n  + D'[\hat z \times \vec v_n ]~.
   \label{Magnus-L} \end{equation}
The forces on the left-hand side of the equation are the {\em 
effective
Magnus force} $\propto \rho_M= \rho_s -D'/\kappa$  and the 
longitudinal friction
force $\propto D$. The forces on the right-hand side are driving 
forces produced
by the superfluid and normal flows. In the theory of 
superconductivity the force
$\rho_s [\vec v_s \times \vec \kappa]$, proportional to the superfluid
velocity $\vec v_s$, is called the Lorentz force. 
The left-hand side of Eq. (\ref{Magnus-L}) presents the response of 
the vortex to these driving forces. The factor $\rho_M$, which determines 
the amplitude of the effective Magnus force on the vortex, is not equal 
to the superfluid density $\rho_s$ in general. In the next sections we
shall consider the contribution to  $D'$ from phonon scattering by a
vortex (the Iordanskii force). The contribution to $D'$ from
the bound states in the vortex core is discussed by Kopnin \cite{KK-S}.

\section{Nonlinear Schr\"odinger equation and two-fluid  
hydrodynamics}
\label{Schrod}

In the Gross--Pitaevskii theory \cite{GP} the ground state and
weakly excited states of a Bose gas are described by the condensate 
wave
function $\psi=a\exp(i\phi)$ which is a solution of the nonlinear
Schr\"odinger equation
\begin{equation}
i\hbar \frac{\partial \psi}{\partial t}= -\frac{\hbar^2}{2m} \nabla^2 
\psi
+V|\psi|^2 \psi~.
   \label{Schr} \end{equation}
Here $V$ is the amplitude of two-particle interaction. The nonlinear
Schr\"odinger equation is  the Euler--Lagrange equation for the
Lagrangian
\begin{equation} 
L={i\hbar \over 2}\left(\psi^*{\partial \psi \over
\partial t} -\psi{\partial \psi^* \over \partial t}\right) 
-{\hbar^2\over
2m}|\nabla \psi|^2 -{V\over 2}|\psi|^4 ~.
   \label{Lagr}   \end{equation}
Noether's theorem yields the momentum conservation law 
\begin{equation} 
\frac{\partial j_i}{\partial t}+\nabla_j \Pi_{ij}
      \end{equation}
where
\begin{equation}
\vec j=-{\partial L\over \partial \dot \psi}\vec \nabla \psi- 
{\partial L\over \partial \dot \psi^*}\dot \psi^*=
-{i\hbar \over 2}\left(\psi^* \vec \nabla \psi -\psi \vec \nabla 
\psi^*\right)
      \end{equation}
is the mass current and
\begin{eqnarray}
\Pi_{ij}=-{\partial L\over \partial \nabla_j \psi} \nabla_i \psi 
-{\partial
L\over \partial \nabla_j \psi^*} \nabla_i \psi^* +L\delta_{ij}
\nonumber \\= {\hbar^2\over 2m} \left(\nabla_i \psi \nabla_j \psi^* 
+\nabla_i
\psi^* \nabla_j 
\psi\right) +\delta_{ij} P
        \label{flux}  \end{eqnarray}
is the momentum-flux tensor. The pressure $P$ in this expression
corresponds to the general thermodynamic definition of the pressure 
via a
functional derivation of the energy with respect to the particle 
density
$n=|\psi|^2$:
\begin{equation}
P=L= n {\delta E \over \delta n} - E 
= n \left[{\partial  E \over \partial n} - \vec \nabla 
 \left({\partial  E \over \partial \vec \nabla n}\right)\right]- E
={V\over 2}|\psi|^4 - {\hbar^2\over 4 m} \nabla^2|\psi|^2~,
  \label{Pqu}    \end{equation}
where 
\begin{equation}
E= {\partial L\over \partial \dot \psi}\dot \psi+ 
{\partial L\over \partial \dot \psi^*}\dot \psi^*-L= 
{\hbar^2\over 2m}|\nabla \psi|^2 +{V\over 2}|\psi|^4   
\end{equation}
is the energy density. But in the hydrodynamic limit (see below) the 
dependence
of the energy on the density gradient is usually neglected. 

Using the Madelung 
transformation \cite{D}, the nonlinear
Schr\"odinger equation (\ref{Schr}) for a complex function may be 
transformed
into two real equations for the liquid density $\rho=ma^2$ and the 
liquid
velocity
$\vec v =(\kappa / 2\pi) \vec \nabla \phi$ where $\kappa =h /m$ is the
circulation quantum. Far from the vortex line these equations are
hydrodynamic equations for an ideal inviscous liquid: 
\begin{equation} 
{\partial \rho \over \partial t} + \vec \nabla(\rho   \vec v) =0~,
     \label{m-cont} \end{equation}
\begin{equation}
{\partial \vec v \over \partial t}+ (\vec v \cdot \vec \nabla)\vec v
= - \vec \nabla \mu~.
    \label{v-Sch} \end{equation}
Here $\mu= Va^2/m$ is the chemical potential. Equation (\ref{flux}) 
becomes
the hydrodynamic  momentum-flux tensor $\Pi_{ij} =P\delta_{ij} + \rho 
v_i(\vec r) v_j(\vec r)$. Thus the hydrodynamics of an
ideal liquid directly follows from the nonlinear Schr\"odinger 
equation.

Suppose that a plane sound wave  propagates in the liquid generating 
the
phase variation $\phi(\vec r,t)=\phi_0 \exp (i\vec k \cdot \vec r - 
i\omega
t)$.
 Then the liquid density and velocity are functions of the time $t$ 
and the
position vector $\vec r$ in the plane $xy$:
\begin{equation}
\rho(\vec r,t) = \rho_0 + \rho_{(1)}(\vec r,t)~,~~~~~
\vec v(\vec r,t) =\vec v_0+ \vec v_{(1)}(\vec r,t)~,
      \label{den-v} \end{equation}
where $\rho_0$ and $\vec v_0$ are the average density and the average 
velocity
in the liquid, $\rho_{(1)}(\vec r,t)$ and  $\vec v_{(1)}(\vec 
r,t)={\kappa
\over 2\pi} \vec \nabla \phi$ are  periodical variations of the 
density and
the velocity due to the sound wave ($\langle \rho_{(1)}
 \rangle =0$, $\langle \vec v_{(1)}\rangle=0$). They should be 
determined from
Eqs. (\ref{m-cont}) and (\ref{v-Sch}) after their linearization. In 
particular,
Eq. (\ref{v-Sch}) gives the relation between the density variation 
and the
phase $\phi$: 
\begin{equation}
\rho_{(1)}= {\rho_0 \over c_s^2} \mu_{(1)}= - {\rho_0 \over c_s^2}
{ \kappa \over 2\pi } \left\{{\partial \phi \over \partial t}+ \vec 
v_0 \cdot
\vec \nabla \phi(\vec r)\right\}~,
     \label{rho-phon0} \end{equation} 
where $c_s=\sqrt{Va^2/m}$ is the sound velocity. Substitution of this 
expression
into Eq. (\ref{m-cont}) yields the sound equation for a moving liquid 
with the
wave spectrum $\omega =c_sk + \vec k \cdot \vec v_0$. 

The sound propagation is accompanied with the transport of mass. This 
is an
effect of the second order with respect to the wave amplitude. The 
total
mass current expanded with respect to the wave amplitude and averaged 
over
time is 
\begin{equation}
\vec j=\rho_0 \vec v_0 + \langle \rho_{(1)} \vec v_{(1)} \rangle +
\langle \rho_{(2)} \rangle \vec v_0 +\rho_0 \langle  \vec v_{(2)}
\rangle~, 
    \label{mass-fl}\end{equation}
where
\begin{equation} 
\langle \rho_{(1)}\vec v_{(1)} \rangle=\vec j^{ph}(\vec p) =\rho_0 
\phi_0^2
{\kappa^2 k \over 8\pi^2 c_s}\vec k =  n(\vec p)  \vec p~.
  \label{mas-fl} \end{equation}
is the mass current and $n(\vec p)$ is the number of phonons  with the
momentum $\vec p =\hbar \vec k$ and the energy $E=\varepsilon (\vec p)
+\vec p \cdot \vec v_0 $. The phonon mass current is the phonon 
momentum
density in the reference frame moving with the average liquid
velocity $\vec v_0$, and $\varepsilon (\vec p) = c_sp $ is the energy 
in the
same reference frame.

Mathematically the second-order corrections to the mass density, 
$\langle
\rho_{(2)} \rangle$, and the average velocity, $\langle  \vec v_{(2)}
\rangle $,  remain undefined, but there are physical constrains to 
specify
them. First of all, we assume that phonon excitations do not change 
the
average mass density, and $\langle \rho_{(2)} \rangle$ must vanish.
As for the second-order correction $\langle  \vec v_{(2)} \rangle $ 
to the
average velocity, it should produce a second-order correction $\langle
\phi_{(2)}
\rangle$ to the phase which is impossible in quantum hydrodynamics. 
The
simplest way to see it is to consider the propagation of phonons in an
annular channel with the periodic boundary conditions. The phase
variation over the channel length is a topological invariant, and weak
excitations (phonons) cannot change it. Therefore $\langle \vec 
v_{(2)}
\rangle $ must vanish. More complicated arguments must be given for an
open geometry, but intuitively it is clear that this basic physical
constrain should not depend on the boundary condition.  

It is important to emphasize a difference between sound waves in a 
liquid and
in an elastic solid. The sound wave in the elastic solid is not
accompanied by real mass transport: all
atoms oscillate near their equilibrium  positions in the crystal 
lattice,
but they cannot move in average if the crystal is fixed at a 
laboratory table. Within our present formalism this means
that the  second-order contribution $\langle 
\vec v_{(2)}\rangle $ to the average velocity must not vanish, but
compensate the  second-order contribution $\langle \rho_{(1)} \vec 
v_{(1)}
\rangle$ to the mass current. Finally a phonon in a solid cannot have 
a real
momentum but only a quasimomentum. The problem of the phonon momentum 
in liquids
and solids has already been discussed a long time
\cite{Peierls} (see also the recent paper by Stone \cite{S-M}), and 
they have
noticed that the sound wave may have a different momentum using Euler 
or
Lagrange variables. In fact, the momentum should not depend on a 
choice of
variables, but only on physical conditions. However, at various 
physical
conditions a proper choice of variables can do an analysis more
straightforward. In solids the Lagrange variables are preferable 
since in
this case the velocity is related to a given particle and coincides 
with  the
{\em center-of-mass} velocity which must not change in average and 
therefore
has no second-order corrections. Using Euler variables in liquids the
{\em average} velocity
$\vec v_0$ relates to a given point in the space and has no 
second-order
corrections due to phonons.

 Expanding the momentum-flux tensor up to the
terms of the second order with respect to the sound wave amplitude one
obtains:
\begin{equation}
\Pi_{ij} =P_0\delta_{ij} + \rho_0  v_{0i} v_{0j}+\Pi^{ph}_{ij}~,
    \label{mom-flux0} \end{equation}
where the second-order phonon contribution is
\begin{equation} 
\Pi^{ph}_{ij} =\langle P_{(2)}
\rangle \delta_{ij} + \langle \rho_{(1)} (v_{(1)})_i \rangle v_{0j} + 
\langle
\rho_{(1)} (v_{(1)})_j  \rangle v_{0i} + \rho_0 \langle (v_{(1)})_i 
(v_{(1)})_j \rangle~.
      \label{mom-ph} \end{equation}
The second-order contribution $P_{(2)}$ to the pressure can be 
obtained from
the Gibbs-Duhem relation $\delta P =\rho \delta \mu$ at
$T=0$ using expansions $\rho = \rho_0+\rho_{(1)}$ and $\mu= \mu_0 +
\mu_{(1)}+\mu_{(2)}$, where $\mu_0$ is the chemical potential without 
the
sound wave. This yields $P_{(2)}=\rho_0
\mu_{(2)} + {\partial \rho \over \partial \mu} {\rho_{(1)}^2 \over 
2}$, where
${\partial \rho \over \partial \mu}=\rho_0/c_s^2$. According to the 
Euler
equation (\ref{Euler}) the second-order contribution  to the chemical
potential is $\mu_{(2)}=-{v_{(1)}^2 \over 2}$. Then  
\begin{equation} 
\langle P_{(2)} \rangle = {c_s^2 \over \rho_0}{\langle
\rho_{(1)}^2 \rangle  \over 2} - \rho_0 {\langle v_{(1)}^2\rangle 
\over 2}~.
     \label{P} \end{equation}
     
In the presence of the phonon distribution the total mass current is
\begin{equation}
\vec j=\rho_0 \vec v_0 +{1 \over h^3} \int  \vec j^{ph}(\vec p)\,d_3 
\vec
p  =\rho_0 \vec v_0 +{1 \over h^3} \int  n(\vec p) \vec p\,d_3\vec p~,
    \label{tot-mass-fl} \end{equation}
In the thermal equilibrium at $T>0$, the
phonon numbers are given by the Planck distribution $n(\vec p)=n_0(E, 
\vec
v_n)$   with the drift velocity $\vec v_n$ of quasiparticles:  
\begin{equation} 
n_0(E, \vec v_n) =\frac{1}{\exp{E(\vec p)- \vec p \cdot \vec v_n 
\over T} -1}
=\frac{1}{\exp{\varepsilon(\vec p)+ \vec p \cdot (\vec v_0 -\vec v_n) 
\over T} -1} ~.
   \label{BE} \end{equation}
Linearizing Eq. (\ref{BE}) with respect to the relative velocity 
$\vec v_0
-\vec v_n$, one obtains from Eq. (\ref{tot-mass-fl}) that
\begin{equation}
\vec j=\rho_0 \vec v_0 +\rho_n (\vec v_n-\vec v_0)~.
    \label{tot-mass-2-fl} \end{equation}
This expression is equivalent to the two-fluid
expression $\vec j = \rho \vec v_s + \rho_n (\vec v_n-\vec v_s) = 
\rho_s \vec
v_s + \rho_n \vec v_n$ assuming that $\rho=\rho_0=\rho_s + \rho_n$, 
$\vec v_0
= \vec v_s$, and the normal density is given by the usual
two-fluid-hydrodynamics expression:
\begin{equation}
\rho_n = -{1 \over 3h^3} \int\frac{\partial n_0(\varepsilon, 
0)}{\partial
E} p^2 \, d_3 \vec p~.
     \label{rho-n} \end{equation} 
In the same manner one can derive the two-fluid momentum-flux tensor 
\cite{PRB7}:
\begin{equation}
\Pi_{ij} = P\delta_{ij} +\rho_s v_{si}v_{sj} +\rho_n  v_{ni}v_{nj}~.
      \label{M-2f} \end{equation}

This analysis demonstrates that two-fluid-hydrodynamics relations can 
be
derived from the hydrodynamics of an ideal liquid in presence of 
thermally
excited sound waves, as was shown by Putterman and Roberts \cite{PR}. 
In
order to obtain a complete system of equations of the two-fluid 
theory, one
should take into account  phonon-phonon interaction, which is 
essential for
the phonon distribution function being close to the equilibrium Planck
distribution. In the two-fluid theory the locally defined superfluid 
and
normal velocities $\vec v_s$ and $\vec v_n$ correspond to the average
velocity of a liquid in a fixed point of the space and to the drift
velocity of the phonon gas respectively. The two-fluid
hydrodynamics is valid only at scales exceeding the phonon mean-free 
path
$l_{ph}$.   

\section{Scattering of phonons by the vortex in hydrodynamics} 
\label{Born}

Phonon scattering by a vortex line was studied beginning from the 
works by
Pitaevskii \cite{PitP} and Fetter \cite{Fet}. We consider a sound wave
propagating in the plane $xy$ normal to a vortex line (the axis $z$). 
In the
linearized hydrodynamic equations of the previous section the fluid 
velocity
$\vec v_0$ should be replaced by the velocity $\vec v_v(\vec r)$ 
around the
vortex line:     
\begin{equation}
{\partial \rho_{(1)} \over \partial t} + \rho_0 \vec \nabla \cdot 
\vec 
v_{(1)} =-\vec v_v \cdot \vec  \nabla \rho_{(1)}~,
     \label{rho} \end{equation}
\begin{equation}
{\partial \vec v_{(1)} \over \partial t}+ {c_s^2 \over \rho_0} 
\vec \nabla \rho_{(1)}=- \left[ (\vec v_v \cdot \vec
\nabla ) \vec v_{(1)} + (\vec v_{(1)} \cdot \vec
\nabla ) \vec v_v  \right]~.
    \label{v-Pit0} \end{equation}
Using the vector identity 
\begin{equation}
(\vec v \cdot \vec \nabla)\vec v = \vec \nabla {v^2 \over 2} - \vec v
\times [\vec \nabla \times \vec v]~
      \label{BAC} \end{equation}
for the velocity $\vec 
v=\vec v_v + \vec v_{(1)}$, Eq. (\ref{v-Pit0}) can be rewritten as
\begin{equation}
{\partial \vec v_{(1)} \over \partial t}+ {c_s^2 \over \rho_0} 
\vec \nabla \rho_{(1)}=-\vec \nabla (\vec v_v \cdot \vec v_{(1)})
+[\vec  v_{(1)} \times \vec \kappa]\delta_2(\vec r)~.
    \label{v-Pit} \end{equation}
The perturbation from the vortex (the right-hand side)  contains a
$\delta$-function because the vortex line is
not at rest when the sound wave propagates past the vortex. In order 
to
weaken the singularity one can introduce the time-dependent vortex 
velocity
$\vec v_v(\vec r,t)$ as a zero-order approximation for the
velocity field \cite{S}. Then $\vec r$ in Eq. (\ref{v-v}) must be
replaced by $\vec r - \vec v_L  t$ and   $\partial \vec v_v/\partial 
t =
-(\vec v_L
\cdot \vec \nabla) \vec v_v = - \vec \nabla (\vec v_L \cdot \vec v_v) 
+[\vec
v_L \times\vec \kappa] \delta_2(\vec r) $. Since there is no
external force on the liquid,  the vortex moves with the velocity in 
the
sound wave: $\vec v_L = \vec v_{(1)}(0,t)$. Now in the linearization
procedure the fluid acceleration in Eq. (\ref{v-Sch}) must be 
presented as
$\partial \vec v/\partial t = \partial \vec v_v/\partial t + \partial 
\vec
v_{(1)}/\partial t$. As a result Eq. (\ref{v-Pit}) is replaced 
by:     
\begin{equation} 
{\partial \vec v_{(1)} \over \partial t}+ {c_s^2 \over
\rho_0}  \vec \nabla \rho_{(1)}= \vec \nabla  [\vec v_v \cdot \vec
v_{(1)}(\vec r)] - \vec \nabla [\vec v_v \cdot \vec v_{(1)}(0)]~.
    \label{v-phon} \end{equation}
Equation (\ref{v-phon}) yields:
\begin{equation}
\rho_{(1)}= - {\rho_0 \over c_s^2}
{ \kappa \over 2\pi } \left\{{\partial \phi \over \partial t}+ \vec 
v_v \cdot
[\vec \nabla \phi(\vec r) -\vec \nabla \phi(0) ]\right\}~.
     \label{rho-phon} \end{equation} 
Substitution of $\rho_{(1)}$ in Eq. (\ref{rho}) yields the linear 
sound
equation for the phase:  
\begin{equation}
\frac{\partial^2 \phi}{\partial t^2} - c_s^2\vec \nabla^2\phi= - 
2\vec v_v
(\vec r) \cdot \vec \nabla \frac{\partial}{\partial t} \left[ 
\phi(\vec r)-
\frac{1}{2} \phi(0) \right]~.
   \label{phi} \end{equation}

In the long-wavelength limit $k \rightarrow 0$ one may use the Born
approximation. The Born perturbation parameter $\kappa k/ c_s$ is on 
the order of
the ratio of the vortex core radius $r_c \sim \kappa/c_s$ to  the 
wavelength
$2\pi/k$. Then after substituting the plane wave into the right-hand 
side of Eq.
(\ref{phi})  the solution of this equation is   
\begin{eqnarray}
\phi =\phi_0 \exp ( - i\omega t)\left\{ \exp (i\vec k \cdot \vec r)
\right. \nonumber \\ \left.
-{i k\over 4c_s} \int d_2\vec r_1 H_0^{(1)} (k|\vec r - \vec r_1|) 
\vec k
\cdot
\vec v_v(\vec r_1) [2\exp(i\vec k \cdot \vec r_1) - 1]\right\}~.
        \label{phi-per} \end{eqnarray}
Here $H_0^{(1)}(z)$ is the zero-order Hankel function of the first 
kind, and
${i\over 4}H_0^{(1)}(k|\vec r - \vec r_1|)$ is the Green function for 
the 2D
wave equation, i. e., satisfies to the equation
\begin{equation}
-(k^2 + \vec \nabla^2)\phi (\vec r)=\delta_2(\vec r- \vec r_1)~.
         \label{Green} \end{equation}

\begin{figure}
\begin{center}
\includegraphics[width=.6\textwidth]{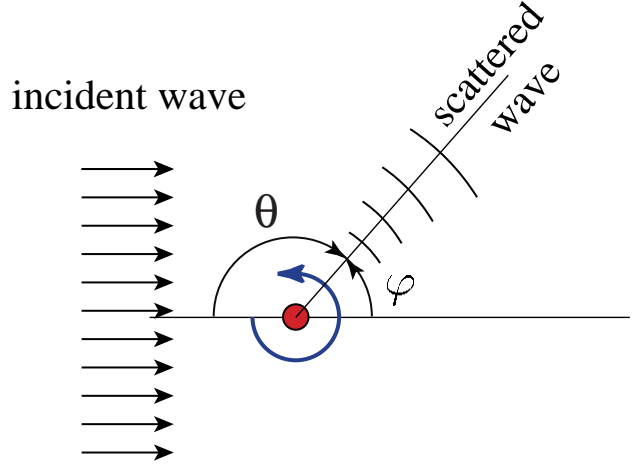}
\end{center}
\caption[]{Scattering of a sound wave by a vortex.}
\label{f2}
\end{figure}

In the standard scattering theory they use
the asymptotic expression for the Hankel function at large values of 
the
argument:
\begin{equation}
\lim_{z \rightarrow \infty} H_0^{(1)}(z) = \sqrt{2\over \pi 
z}e^{i(z-\pi/4)}~.
     \label{asym} \end{equation}
If the perturbation is confined to a vicinity
of the line, then $r_1 \ll r$ and
\begin{equation}
|\vec r - \vec r_1| \approx r - \frac{(\vec r_1 \cdot \vec r)}{r}~.
          \label{r-r-1} \end{equation}
After integration in Eq. (\ref{phi-per}) the wave
at  $kr \gg 1$ becomes a superposition of the incident
plane wave $\propto \exp (i\vec k \cdot \vec r)$ and the scattered 
wave
$\propto \exp (ik r)$:  
\begin{equation}
\phi = \phi_0 \exp ( - i\omega t)\left[ \exp (i\vec k \cdot \vec r) +
\frac{ia(\varphi)}{\sqrt{r}} \exp(ikr)\right]~.
       \label{asymptot} \end{equation}
Here $a(\varphi)$ is the scattering amplitude which is a function of 
the angle
$\varphi$ between the initial wave vector $\vec k$ and the wave 
vector $\vec
k' =k\vec r/r$ after scattering (see Fig. \ref{f2}). In the Born 
approximation
(see the paper \cite{PRB7} and references therein for more details) 
\begin{equation}
a(\varphi) =\sqrt{{k \over 2\pi}}\frac{1}{c_s}e^{i{\pi \over 4}} [\hat
\kappa \times \vec k']\cdot \vec k {1 \over q^2}\left(1 -{q^2\over 
2k^2}
\right)  = -{1 \over 2}\sqrt{k\over 2\pi}\frac{\kappa }{c_s}e^{i{\pi 
\over 4}}
\frac{\sin \varphi \cos \varphi}{1-\cos \varphi}~,
     \label{a-vort} \end{equation}
where $\vec q =\vec k - \vec k'$ is the momentum transferred by the 
scattered
phonon to the vortex, and $q^2=2k^2(1-\cos \varphi)$.

Thus the vortex is a line defect which scatters a sound wave.
Scattering produces a force on the defect (vortex). If the
perturbation by the line defect is 
 confined to a finite vicinity of the line,  the force 
\begin{equation}
\vec F^{ph} =  \sigma_{\parallel} c_s \vec j^{ph} - 
\sigma_{\perp} c_s [\hat z \times \vec j^{ph}]
       \label{force} \end{equation}
is determined by two effective cross-sections \cite{S,PRB7}: the 
transport
cross-section for the dissipative force,  
\begin{equation} 
\sigma_\parallel = \int \sigma (\varphi)(1-
\cos \varphi) d\varphi~,
    \label{sig-C} \end{equation}
and the transverse cross-section for the transverse force,
\begin{equation}
\sigma_\perp = \int \sigma (\varphi) \sin \varphi d\varphi~.
    \label{sig-S} \end{equation}
The differential cross-section $\sigma (\varphi)=|a(\varphi)|^2$ with
$a(\varphi)$ from  Eq. (\ref{a-vort})  is
quadratic in the circulation $\kappa$ and even in $\varphi$. 
Therefore in the
Born approximation the transverse cross-section $\sigma_\perp$ 
vanishes.

However, due to a very slow decrease of the velocity $v_v \propto 
1/r$ far
from the vortex, the scattering amplitude is divergent at small 
scattering
angles $\varphi \rightarrow 0$:
\begin{equation}
 \lim_{\varphi \rightarrow 0} a(\varphi) =-
\sqrt{{k \over 2\pi}}\frac{\kappa }{c_s}e^{i{\pi \over 
4}}\frac{1}{\varphi}~.
     \label{a-sm-phi} \end{equation}
This divergence is integrable in the integral for the transport 
cross-section,
Eq. (\ref{sig-C}), and its calculation is reliable. Contrary to it, 
the integrand in Eq. (\ref{sig-S}) for the
transverse cross-section has a pole at $\varphi =0$. The principal 
value of the
integral vanishes, but there is no justification for the choice of 
the principal
value, and the contribution of the small angles requires an 
additional analysis. 

At small scattering angles $\varphi \ll 1/\sqrt{kr}$ the asymptotic
expansion Eq. (\ref{asymptot}) is invalid, and one cannot use the 
differential
cross-section or the scattering amplitude for description of the 
small-angle
scattering \cite{I,S,Berry,Sh}. Meanwhile, the small-angle behavior 
is crucial for
the transverse force as demonstrated below. The accurate calculation 
of the
integral in Eq. (\ref{phi-per}) for small angles (see Ref.
\cite{S} and  Appendix B in Ref. \cite{PRB7}) yields that at 
$\varphi \ll 1$  
\begin{equation} 
\phi =\phi_0 \exp ( - i\omega t+i\vec k \cdot \vec r)
\left[1+\frac{i\kappa k}{2c_s} \Phi\left(\varphi \sqrt{kr \over
2i}\right)\right]~.
        \label{phi-small} \end{equation}
 Using an asymptotic expression for the error integral  
\begin{equation}
\Phi(z)={2 \over \sqrt{\pi}} \int \limits_0^z e^{-t^2} dt 
\longrightarrow  
{z \over |z|}- {1 \over \sqrt{ \pi} z} \exp
(-z^2)  
     \label{Phi} \end{equation}
 at $|z| \rightarrow \infty$, one obtains  for angles $ 1 \gg \varphi 
\gg
1/\sqrt{kr}$:
\begin{equation} 
\phi =\phi_0 \exp ( - i\omega t)\left[ \exp(i\vec k \cdot \vec r)
\left(1+\frac{i\kappa k}{2c_s} {\varphi \over |\varphi|}  \right)- 
\frac{i\kappa
}{c_s}\sqrt{k \over 2\pi r} {1 \over \varphi} \exp\left(ikr + i{\pi 
\over
4}\right) \right] ~.
        \label{phi-clas} \end{equation}
The second term in square brackets coincides with scattering wave at 
small
angles   $\varphi \ll1$ with the amplitude given by Eq.
(\ref{a-sm-phi}). But now one can see that the standard scattering 
theory
misses to reveal an important non-analytical correction to the 
incident
plane wave, which changes a sign when the scattering angle 
$\varphi$ crosses zero. Its physical meaning is discussed in the next 
section.

\section{The Iordanskii force and the Aharonov--Bohm effect} 
\label{AB-I}

The analogy between the phonon scattering by a vortex and the 
Aharonov--Bohm
effect for electrons scattered by a magnetic-flux tube becomes 
evident if one
rewrites the sound equation (\ref{phi}) in presence of the vortex as
\begin{equation} 
k^2 \phi- \left(-i\vec \nabla + {k \over c_s} \vec v_v\right)
^2\phi= 0~.
   \label{phon-vor} \end{equation}
It differs from the sound equation  (\ref{phi}) 
by the term of the second order in $v_v \propto \kappa$ and by 
absence of the 
contribution from the vortex-line motion [the term $\propto \phi(0)$ 
on the
right-hand side of  Eq. (\ref{phi})]. This difference is  unimportant 
for the
calculation of the transverse force, which is linear in $\kappa$. On 
the other
hand, the stationary Schr\"odinger equation for an electron in 
presence of
the magnetic flux $\Phi$ confined to a thin tube (the Aharonov--Bohm 
effect
\cite{AB}) is:  
\begin{equation} E\psi(\vec r) = {1 \over 2m}\left(-i\hbar  \vec
\nabla -{e \over c}\vec A \right)^2 \psi (\vec r)~.
      \label{AB} \end{equation}
Here $\psi$ is the electron wave function with energy $E$ and the
electromagnetic vector potential is connected with the magnetic flux 
$\Phi$ by the
relation similar to that for the velocity $\vec v_v$ around the 
vortex line
[Eq. (\ref{v-v})]: 
\begin{equation}
\vec A =\Phi \frac{[\hat z \times \vec r]}{2\pi r^2}~.
         \label{A} \end{equation}

Let us consider the quasiclassical solution of the sound equation:
\begin{equation}
\phi =\phi_0 \exp \left( - i\omega t+i\vec k \cdot \vec r+ {i\delta S
\over \hbar}\right)
 =\phi_0 \exp ( - i\omega t+i\vec k \cdot \vec r)
\left(1+ \frac{i\kappa k}{2\pi c_s} \theta    \right)  ~,
        \label{quasiclas} \end{equation}
where $\delta S=-(\hbar k/c_s) \int ^{\vec r} \vec v_v \cdot d\vec l 
={\hbar
\theta
\kappa k/2\pi c_s}
$ is the variation of the action due to interaction with the circular
velocity from the vortex along quasiclassical trajectories. The angle
$\theta$ is an  azimuth angle for the position vector $\vec r$
measured from the direction opposite to the wave vector $\vec k$ (see 
Fig.
\ref{f2}). This choice provides that the quasiclassical correction 
vanishes for the
incident wave far from the vortex. One can check directly that Eq.
(\ref{quasiclas}) satisfies the sound equation (\ref{phi}) in the  
first
order of the parameter $\kappa k/c_s$. For the Aharonov--Bohm effect 
the
phase $\delta S/\hbar$ arises from the electromagnetic vector 
potential:
$\delta S = -( e/c) \int ^{\vec r} \vec A \cdot d\vec l$.

The velocity generated by the sound wave around the vortex is
\begin{equation}
\vec v_{(1)}= {\kappa\over 2\pi}\vec \nabla \phi =
{\kappa\over 2\pi} \phi_0 \exp ( - i\omega
t+i\vec k \cdot \vec r)
\left(i\vec k - \frac{i k}{ c_s} \vec v_v   \right)  ~.
        \label{QCv} \end{equation}
From Eqs. (\ref{rho-phon}) and (\ref{QCv}) we can obtain the  phonon 
mass 
current with first-order corrections in $\vec 
v_{v}$. 
\begin{equation} 
\vec j_{ph}=\langle \rho_{(1)} \vec v_{(1)} \rangle ={1\over 
8\pi^3}\int
d\vec k\,n(\hbar
\vec k)\hbar
\left[\vec k - 
\frac{k}{ c_s} \vec v_{v}(\vec r)
 - (\vec k \cdot \vec v_{v}(\vec r) ){\vec k \over c_s k} \right]~.
  \label{mas-vor} \end{equation} 
Due to the last term in this expression, the phonon mass current is 
not 
curl-free. 

According to  Eq. (\ref{quasiclas}) the phase $\phi$ is multivalued, 
and one must choose a cut for an angle $\theta$ at the direction $\vec
k$,  where $\theta  =\pm \pi$ .
The jump of the phase on the cut line behind the vortex is a
manifestation of the Aharonov--Bohm effect
\cite{AB}: the sound wave after its interaction with the vortex
has different phases on the left and on the right of the vortex 
line. This
results in an interference \cite{S,PRB7}. 

\begin{figure}
\begin{center}
\includegraphics[width=.9\textwidth]{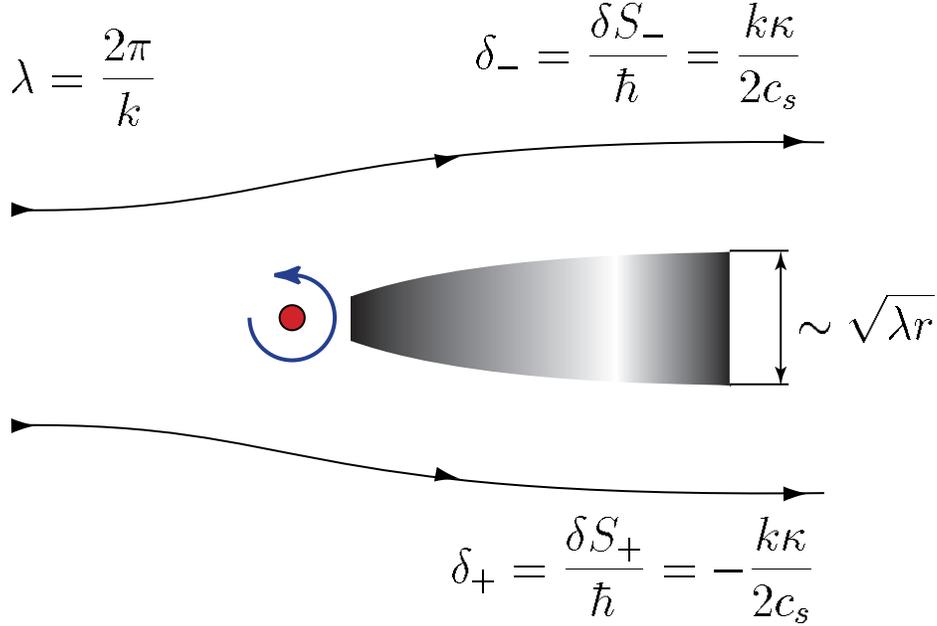}
\end{center}
\caption[]{Aharonov--Bohm interference.}
\label{f4}
\end{figure}

In the interference region the quasiclassical solution is invalid and 
must be
replaced by Eq.  (\ref{phi-small}). The width of the interference 
region is 
 $d_{int}\sim \sqrt{r /k}$. Here $r$ is the 
distance from the vortex line. The interference region corresponds to 
very
small scattering angles $\sim d_{int}/r=1/\sqrt{kr}$.

Now we are ready to consider the momentum balance using the condition 
that
$\int dS_j \Pi_{\perp j}=0$ where subscript $\perp$ is for a component
normal to the wave vector $\vec k$ of the incident wave.  The total
momentum-flux tensor can be obtained from Eqs. (\ref{mom-flux0}) and
(\ref{mom-ph}) assuming $\vec v_0 (\vec r)=
\vec v_v(\vec r) + \vec v_s$ and neglecting some unimportant terms:
\begin{eqnarray}
\Pi_{ij}= -\rho_0 (\vec v_s -\vec v_L)\cdot \vec v_v 
\delta_{ij} + \rho_0  v_{0i} v_{0j}
\nonumber \\ + \langle \rho_{(1)} (v_{(1)})_i \rangle
v_{vj} + \langle \rho_{(1)} (v_{(1)})_j  \rangle v_{vi} + \rho_0 
\langle
(v_{(1)})_i  (v_{(1)})_j \rangle~.
   \label{fluxF} \end{eqnarray}
The first two terms in this expression yield the momentum flux without
phonons, which gives the Magnus force
for a liquid with the density $\rho_0$ and the velocity $\vec v_s$. 
The term
$\propto v_{vj}$ does not contribute to the momentum flux through a
cylindrical surface around the vortex, since the velocity  $\vec v_v$ 
is
tangent to this surface. The term $ v_{v_i}\langle  \rho_{(1)} 
(v_{(1)})_j \rangle$, in which  the mass current  $\langle  
\rho_{(1)} 
(v_{(1)})_j \rangle$ is given  Eq.~(\ref{mas-fl}) for the plane wave 
in
absence of the vortex, gives a
contribution to the momentum flux, which  exactly cancels the
contribution of the term $\rho_0 \langle (v_{(1)})_i  (v_{(1)})_j 
\rangle$ 
outside the interference region, where  the velocity is given by Eq.
(\ref{QCv}). Finally only the interference
region contributes the momentum flux $\int dS_j \Pi_{\perp j}$.

In the interference region the velocity is obtained by taking the 
gradient of
the phase given by Eq. (\ref{phi-small}). Its component normal to 
$\vec k$, 
\begin{equation}
v_{(1)\perp}= {\kappa \over 2\pi r} {\partial \phi \over \partial 
\varphi}
= \phi_0 \exp(-i\omega t + i\vec k \cdot \vec r) 
{i\kappa^2 k \over 4\pi c_s r} {\partial \Phi\left(\varphi \sqrt{kr /
2i}\right) \over \partial \varphi} ~,
       \label{v-trans} \end{equation}
determines the interference contribution to the transverse force:
\begin{eqnarray}
\int dS_j \rho_0 \langle (v_{(1)})_\perp (v_{(1)})_j \rangle
= \int \rho_0 \langle (v_{(1)})_\perp (v_{(1)})_r \rangle rd\varphi
\nonumber \\
={1 \over 8\pi^2} \rho_0 \phi_0^2 \frac{\kappa^2 k}{\hbar }
\left[\delta S_- -\delta S_+\right] = 
{1 \over 8\pi^2} \rho_0 \phi_0^2 \frac{\kappa^3 k^2}{c_s}=\kappa 
j^{ph}~,
    \label{f-inter} \end{eqnarray}
where $\delta S_\pm$ are the action variations at $\theta \rightarrow 
\mp \pi$. Thus the interference region, which corresponds to  an
infinitesimally small angle interval, yields a finite contribution to 
the
transverse force, which one could not obtain from the standard 
scattering
theory using the differential cross-section. In fact the details of 
the
solution in the  interference region are not essential: only a jump 
of the phase
across the  interference region is of importance.

Finally the momentum balance condition
$\int dS_j \Pi_{\perp j}=0$  yields the relation
\begin{equation}
\rho_0 [(\vec v_L - \vec v_{s}) \times \vec \kappa] -
[\vec j^{ph}(\vec p) \times \vec \kappa]=0~. 
   \label{Magnus-ph} \end{equation}
The second vector product on the left-hand side is a transverse force
which corresponds to  the transverse cross-section [see Eq. 
(\ref{sig-S})]
\begin{equation}
\sigma_\perp= \frac{\delta S_- -\delta S_+}{\hbar k} 
   \label{sig-Act} \end{equation}
equal to $\kappa /c_s$. For the
Planck distribution of phonons, $\vec j^{ph}$ must be replaced by
$\rho_n (\vec v_n - \vec v_s)$:
\begin{equation}
\rho_s [(\vec v_L - \vec v_{s}) \times \vec \kappa] + \rho_n
[(\vec v_L - \vec v_n) \times \vec \kappa]=0~. 
   \label{Iordan} \end{equation}
The term $\propto (\vec v_L - \vec v_n)$ is the Iordanskii force, 
which
corresponds to $D'=-\kappa \rho_n$ in Eq. (\ref{F}). The
longitudinal force $\propto D$ is not present in Eq. (\ref{Iordan}) 
since
we ignored terms of the second order in $\kappa$ in order to simplify
our derivation. According to Eq. (\ref{Iordan}) the vortex moves with 
the center-of-mass
velocity $\vec v={\rho_s \over \rho}\vec v_s +  {\rho_n \over 
\rho}\vec v_n$.

Our scattering analysis was done in the coordinate frame where the 
vortex is
at rest and we neglected the relative superfluid velocity $\vec v_s - 
\vec
v_L$ with respect to the vortex. If the velocity $\vec v_s - \vec
v_L$ is high it can affect the value of the effective cross-section. 
But
since the phonon momentum is linear with respect to $\vec v_n - \vec 
v_s$,
the dependence of the cross-section on $\vec v_s - \vec v_L$ is a 
nonlinear
effect. Thus our momentum balance took into account all
effects linear in the superfluid velocity $\vec v_s$. This is 
confirmed by a
more elaborate analysis of Stone \cite{stone}.

\section{Partial-wave analysis and the Aharonov--Bohm effect} 
\label{AB-PW}

Studying interaction of phonons with a vortex  we solved the sound
equation using the perturbation theory. It is completely justified 
because the
parameter of the perturbation theory $\kappa k/c_s$ is the ratio 
between
the vortex core and the phonon wavelength, which is always small for 
phonons. But
in  the Aharonov--Bohm problem for electrons the corresponding 
parameter is
$\gamma =
\Phi/\Phi_1$, where  $\Phi_1 =hc/e$ is the magnetic-flux quantum for 
one electron
(two times larger than  the magnetic-flux quantum $\Phi_0=hc/2e$  for 
a Cooper
pair). This parameter can be arbitrary large, and the perturbation 
theory is not
enough to describe an expected periodic dependence on $\gamma$. On 
the other hand,
there is an exact solution of the Aharonov--Bohm problem for electrons 
obtained by
the partial-wave expansion \cite{AB}, and it will be shown now how to 
derive the
transverse force from this solution. Another derivation of the force 
on the 
 Aharonov--Bohm flux tube using the wave-packet presentation is 
discussed by
Shelankov \cite{ShelB}.  

We look for a solution of Eq. (\ref{AB}) as a superposition of the 
partial
cylindrical waves $\psi= \sum \limits_l \psi_l(r)\exp(il\varphi)$.
Partial-wave amplitudes $\psi_l$ should satisfy equations in the 
cylindrical system
of coordinates $(r,\varphi)$:
\begin{equation}
\frac{d^2\psi_l}{dr^2} + {1 \over r} \frac{d\psi_l}{dr} 
-\frac{(l-\gamma)^2}{r^2} \psi_l + k^2 \psi_l=0~. 
   \label{AB-l} \end{equation}
Here $k$ is the wave number of the electron far from the vortex so 
that
$E=\hbar^2k^2 /2m$. We need a solution of Eq. (\ref{AB-l}), which  at 
large
distances has an asymptotic behavior given by  Eq. (\ref{asymptot}):
\begin{equation}
\psi_l =   \sqrt{2 n \over \pi kr} \exp\left[ i{\pi \over 2} 
l+i\delta_l
\right]  \cos(kr - {\pi \over 2}  l +\delta_l -{\pi \over 4})~,
  \label{PW-ampl} \end{equation}
where  $n$ is the particle density and the partial-wave phase shifts
are 
\begin{equation}
\delta_l =(l-|l-\gamma|)\pi/2~.
  \label{ph-sh-AB}\end{equation}
For $\delta_l=0$ Eq. (\ref{PW-ampl}) yields the partial-wave 
amplitudes of the
incident plane wave at large distances $r$. But for nonzero 
$\delta_l$ there is
also the scattered wave in Eq. (\ref{asymptot}) with the
scattering amplitude
\begin{equation}
a(\varphi) =\sqrt{1 \over 2\pi k}\exp\left(i{\pi \over 4}\right)
\sum \limits_l [1 - \exp(2i\delta_l)] \exp(il\varphi)~.
      \label{scat-phase}  \end{equation}
      
The transverse force is determined by the  transverse cross-section
\begin{eqnarray}
\sigma_\perp = \int |a(\varphi)|^2 \sin \varphi d\varphi \nonumber \\
={1 \over 2i k}\left\{\sum \limits_l e^{i2\delta_{l-1}}-\sum \limits_l
e^{i2\delta_{l+1}} + \sum \limits_l e^{i2(\delta_{l+1} - \delta_l)}
-\sum \limits_l e^{i2(\delta_{l-1} - \delta_l)}\right\}~.
    \label{l-sum}   \end{eqnarray}
Shifting the number $l$ by 2 in the first sum and by one in the 
fourth sum one
obtains the expression  for the transverse cross-section   in the 
partial-wave
method derived long ago  by Cleary \cite{Clea}:
\begin{equation}
\sigma_\perp = \int |a(\varphi)|^2 \sin \varphi d\varphi
={1 \over k}\sum \limits_l \sin2(\delta_l - \delta_{l+1})~.
    \label{sig-S-pw} \end{equation}
Using the phase shift values for the Aharonov--Bohm solution, Eq.
(\ref{ph-sh-AB}), the transverse cross-section is
\begin{equation}
\sigma_\perp=-{1 \over k}\sin 2\pi \gamma~.
         \label{sig-S-AB} \end{equation}
However, the shift of $l$ in the first sum of Eq. (\ref{l-sum}) is
not an innocent operation because of divergence of the first and 
second sum at
$l\rightarrow \pm \infty$. The derivation of Clearly's formula 
(\ref{sig-S-pw})
assumes that the first and the second divergent sums in Eq. 
(\ref{l-sum}) should
exactly cancel after the shift of $l$. But if one does not shift 
$l$, the difference of the first and the second sum is
finite. Moreover, this difference cancels the contribution of the 
third and the
fourth sum, and $\sigma_\perp $ vanishes in the first order with 
respect to the phase shifts
$\delta_l$.

Ambiguity in the calculation of the partial-wave sum for the 
transverse
cross-section is another manifestation of the small-scattering-angle 
problem in
the configurational space: the standard scattering theory does not
provide a recipe how to treat a singularity at zero scattering angle. 
The
zero-angle singularity is responsible for a divergent partial-wave 
series. The way
to avoid the ambiguity is similar to that in the configurational 
space: one should
not use the concept of the scattering amplitude for calculation of 
the transverse
force. 

We must analyze the momentum balance. The momentum-flux tensor for the
electron Schr\"odinger equation (\ref{AB}) is 
\begin{eqnarray}
\Pi_{ij}= {1\over 2 m}\mbox{Re}\left\{
\left(-i\hbar  
\nabla_i -{e \over c} A_i\right) \psi \left(i\hbar  
\nabla_i -{e \over c} A_i\right) \psi^* \right.\nonumber \\ \left. 
- \psi^*  \left(-i\hbar  
\nabla_i -{e \over c} A_i\right)\left(-i\hbar  
\nabla_j -{e \over c} A_j\right) \nabla_j \psi \right\}~.
   \label{fluxAB}   \end{eqnarray}

If the axis $x$ is directed along the wave vector $\vec k$ of the incident
wave, the transverse force is determined by the momentum-flux tensor
component $\Pi_{yr}$, where $r$ is the radial coordinate in the 
cylindrical coordinate system $r=\sqrt{x^2+y^2}$ and $\varphi = \arctan (y/x)$. 
In Eq. (\ref{fluxAB}) the terms $A$ and $\partial /r \partial \varphi$, 
which are inversely proportional to $r$, are not important at large $r$. Then
\begin{equation}
\Pi_{yr}={\hbar^{2}\over 2m}\sum_{l'} \sum_l \mbox{Re}\left\{\sin 
\varphi \left[
{\partial \psi^*_{l'}\over \partial r} {\partial \psi_{l}\over 
\partial r} -
\psi_l^* {\partial^2 \psi^*_{l}\over \partial
r^2}\right]e^{i(l-l')\varphi}\right\}~,
     \end{equation}
where $\psi_l$ are given by Eq. (\ref{PW-ampl}).
Finally the transverse force is
\begin{equation}
F_\perp=\oint \Pi_{yr}rd\varphi =\pi r {\hbar^{2}\over 
m}\mbox{Im}\sum_l  \left\{
{\partial \psi^*_{l+1}\over \partial r} {\partial \psi_{l}\over 
\partial r}  -
\psi_l^* {\partial^2 \psi^*_{l+1}\over \partial r^2} \right\}~.
      \end{equation}
Here we also made a shift of $l$ in some sums, but it is not dangerous
because the terms of these sums decrease in the limits $l\rightarrow 
\pm
\infty$. 

Inserting the partial-wave amplitudes and their radial derivatives 
into the
expression for the force we obtain:
\begin{equation}
F_\perp =-k n {\hbar^{2}\over m}\sum_l  \sin 2 (\delta_{l+1}-\delta_l)
=- jv \sigma_\perp~,
   \end{equation} 
where $j =\hbar k n$ and $v=\hbar k/m$ are the momentum density and 
the
velocity in the incident plane wave, and $\sigma_\perp$ is the 
effective
cross-section Eq. (\ref{sig-S-pw}) derived by Cleary \cite{Clea}. 
Thus there is a
well-defined effective transverse cross-section, despite that we 
cannot use
directly its expression (\ref{sig-S}) via the differential 
cross-section. But
formally we can ``repair'' this expression by a recipe how to treat 
the
singularity at small angles: we should add to the differential 
cross-section a
singular term proportional to the derivative of $\delta(\varphi)$, 
and to take the
principle value of the integral over the rest part of the differential
cross-section. 

One can obtain  the transverse cross-section for phonons directly from
Eq. (\ref{sig-S-pw}) assuming that $\sin 2\pi \gamma \approx 
2\pi\gamma=-\kappa k/
c_s$. However, it is useful to follow the connection between Cleary's 
formula
(\ref{sig-S-pw}) and Eq. (\ref{sig-Act}) obtained from the 
quasiclassical
solution. In the classical limit the partial wave $l$ corresponds to 
the
quasiclassical trajectory with the impact parameter $b=l/k=\hbar l/p$ 
and the
small scattering angle is $\varphi=- d\delta_l/dl=-(1/2\hbar k) 
d\delta S(b)/ db$,
where
$\delta S(b)$ is the action variation along the trajectory with 
impact parameter
$b$. Finally replacing the sum by an integral Eq. (\ref{sig-S-pw}) 
can be rewritten
as 
\begin{equation}
\sigma_\perp =-{2 \over k}\int_{-\infty}^\infty dl {d\delta_l \over 
dl}
=-{1 \over \hbar k} \int_{-\infty}^\infty db{d\delta S(b) \over db} ~.
    \label{sig-S-n} \end{equation}
which yields  (\ref{sig-Act}) since $\delta S_\pm= \delta S(\pm 
\infty)$. 
Strictly speaking Eq. (\ref{sig-S-n}) is valid if $\delta S(b)$ is a
continuous function. This is the case for rotons which are scattered
quasiclassically
\cite{S,RMP,PRB7}. But it yields a correct answer even for phonons 
despite
$\delta S(b)$ has a jump at $b \sim 0$. Thus even though  phonon 
scattering
cannot be described in the quasiclassical approximation (it yields 
the scattering
angle $\varphi=- (1/\hbar k) d\delta S(b)/ db \approx 0$ at $b \neq 
0$), the
quasiclassical expression (\ref{sig-S-n}) gives a correct phonon 
transverse
cross-section. 

\section{Momentum balance in the two-fluid hydrodynamics} \label{2D}

Up to now we analyzed spatial scales much less than the 
mean-free-path $l_{ph}$
of phonons (ballistic region). Sound waves (phonons) interacted with 
the velocity
field generated by a vortex, but phonon-phonon interaction was 
neglected.
Now we shall see what is going on at scales
much larger than $l_{ph}$, where the two-fluid hydrodynamics is valid.

Scattering of phonons by the vortex  in the ballistic region produced 
a force
$\vec F$ on the vortex, and according to the third Newton law a force 
$-\vec F$ on
the phonons (the normal fluid) should also arise. A momentum transfer 
from a
phonon to a vortex  takes place at distances of the order of the 
phonon wavelength
$\lambda=2\pi/k$, which is much less than the mean-free path $l_{ph}$. 
This means
that at hydrodynamic scales the force on the normal fluid is a
$\delta$-function force concentrated along the vortex line. The 
response of the
normal fluid to this force is described by  the  Navier-Stokes 
equation with the
dynamic  viscosity 
$\eta_n$:
\begin{equation} 
{\partial \vec v_{n}\over \partial t}+(\vec v_{n} \cdot \vec
\nabla) \vec v_{n} = \nu_n \Delta \vec v_{n} -{\vec \nabla P \over 
\rho}-{\rho_s
S \over \rho_n \rho}\vec \nabla T~,
        \label{NS}  \end{equation}  
where $\nu_n=\eta_n /\rho_n$ is the kinematic viscosity, $S$ is the 
entropy per
unit volume, and $T$ is the temperature. Our problem is similar to the 
Stokes
problem of a cylinder moving through a viscous liquid \cite{Magnus}. 
Neglecting
the nonlinear inertial (convection) term $(\vec v_{n} \cdot \vec 
\nabla) \vec
v_{n}$ in the Navier-Stokes equation, the $\delta$-function force 
produces  a
divergent logarithmic velocity field (the Stokes paradox 
\cite{Magnus}):
\begin{equation}
\vec v_{n}(\vec r) = \vec v_{n}+\frac{\vec F}{4\pi \eta_n } \ln{r 
\over
l_{ph}}~,
        \label{Os}  \end{equation} 
where $\vec v_{n}$ is the normal velocity at a distance of order $r 
\sim  l_{ph}$,
which  separates the ballistic and the hydrodynamic regions. In this 
expression
the mean-free path $l_{ph}$ replaces the cylinder radius of the 
classical Stokes
problem in the argument of the logarithm. Such a choice of the lower 
cut-off
assumes that the quasiparticle flux on the vortex is entirely 
determined by the
equilibrium Planck distribution at the border between the ballistic 
and
the two-fluid-hydrodynamics region at $r \sim l_{ph}$ \cite{Low}. 

\begin{figure}
\begin{center}
\includegraphics[width=.95\textwidth]{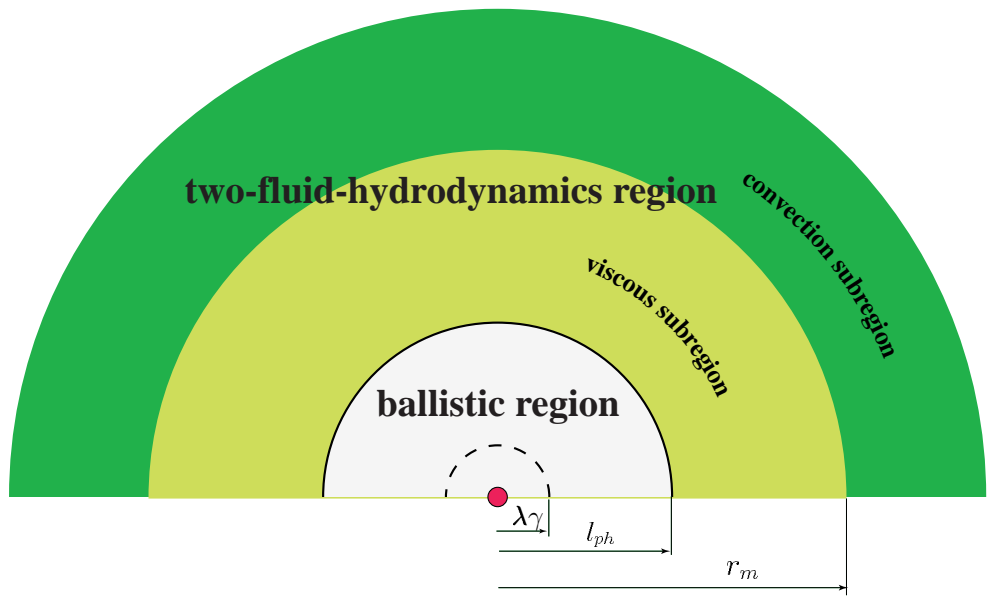}
\end{center}
\caption[]{Relevant scales: the phonon wavelength $\lambda$, the 
phonon mean free
path $l_{ph}$, and Oseen's length $r_m\sim \nu_n /|\vec v_n- \vec 
v_L|$}
\label{f5}
\end{figure}

However small the relative normal
velocity $\vec v_n- \vec v_L$ could be, at  distance of the order or 
larger than
$r_m\sim
\nu_n /|\vec v_n- \vec v_L|$ the nonlinear convection term becomes 
important and stops
a logarithmic growth of the normal velocity. Due to the force $\vec 
F$ the
normal velocities $\vec v_{n\infty}$ and $\vec v_{n}$ at large ($r 
\sim r_m$) and
small ($r \sim l_{ph}$) distances from the vortex line are different 
(the viscous
drag \cite{HV}):
\begin{equation}
   \vec F = \frac{4\pi \eta_n }{\ln(r_m/l_{ph})}(\vec 
v_{n\infty}-\vec v_n)~.
         \label{f-v}  \end{equation}
At very large distances  $r \gg r_m$ the nonlinear convection term is 
more
important than the viscous term. Thus the scale $r_m$, which we shall 
call
Oseen's length, divides the hydrodynamic region onto the viscous and 
convection
subregions \cite{Upp}. All relevant scales are shown in Fig \ref{f5}. 
For a
longitudinal force the solution of the Navier-Stokes equation valid 
for both the
viscous and convection subregions was obtained by Oseen long ago (see 
Ref.
\cite{Magnus}). 

The force $- \vec F$ is transmitted to large distances by a constant 
momentum flux.
In the viscous subregion momentum transport in the normal fluid 
occurs via
viscosity: $F_{i}=-\oint \tau_{ij} dS_{j}$, where $ 
\tau_{ij}=-\eta_n\left(\nabla_i
v_{nj} +\nabla_j v_{ni}\right)$ is the viscous stress tensor. On the 
other hand,
the total momentum flux for the whole liquid should vanish: $\oint
\Pi_{ij}dS_j +\oint \tau_{ij} dS_{j}=0$, where $\Pi_{ij}$ is given by 
Eq.
(\ref{M-2f}).  The normal velocity field does not contain the circular
velocity $\vec v_{v}$, and there is no normal circulation in the 
viscous subregion.
Therefore the momentum flux $\oint \Pi_{ij}dS_j$ yields  the 
superfluid Magnus
force, i.e., the momentum balance in the viscous subregion confirms 
that the force
$\vec F$  satisfies Eq. (\ref{Magnus-S}).  

In the convection region the viscosity becomes ineffective and the 
momentum
conservation gives again $\oint \Pi_{ij}dS_j=0$, like in the 
ballistic region. The
superfluid part of the momentum flux is related to the superfluid 
Magnus force
$\propto \rho_s$ and the normal part of the flux should transmit the 
same force $\vec
F$. The relation between the force $\vec F$ and the relative normal  
velocity 
$ \vec v_{n\infty}-\vec v_L $ in the convection subregion can be 
derived from Eqs.
(\ref{F}) and (\ref{f-v}):  
\begin{equation}
\vec F = -\tilde D(\vec v_L - \vec v_{n\infty}) - \tilde
D'[\hat z \times (\vec v_L - \vec v_{n\infty})]~,
     \label{Fcon} \end{equation} 
where the parameters $\tilde D$ and $\tilde D'$ are connected with 
$D$ and
$D'$ in Eq. (\ref{F}) by a complex relation
\begin{equation}
{1\over \tilde D + i \tilde D'}={\ln(r_m/l_{ph})\over 4\pi \eta_n}
+ {1\over D+i D'}~.
   \label{DD}  \end{equation}
Since in the convection region the viscosity is ineffective and the 
normal fluid
behaves as an ideal one, the only way to transmit the transverse 
component of the
force $\vec F$ to infinity is to create a circulation of the normal 
velocity \cite{Pit}. 
 Because of the viscous drag separation on the longitudinal and 
the transverse
force should be done with respect to the normal velocity
$\vec v_{n\infty}-\vec v_L$, but not $\vec v_{n}-\vec v_L$. Thus the 
normal
circulation is determined by  $\tilde D'$, but not $D'$:
\begin{equation}
\kappa_n= \oint d\vec l \cdot \vec v_n=- {\tilde D' \over \rho_n}
= \frac{\kappa}{1+
\left[{\kappa \rho_n\ln(r_m/l_{ph})
\over 4\pi \eta_n }\right]^2}~,
     \label{circ} \end{equation} 
where we neglected the longitudinal component $\propto D$ in the 
ballistic region and
used the relation $D'=-\kappa \rho_n$ for the Iordanskii force.  

In the convection subregion more problematic is to transmit not the
transverse, but the longitudinal component of the force. A longitudinal
force is impossible if viscosity is neglected completely: a body moving
through an ideal  liquid does
not produce any dissipative force (d'Alembert's paradox \cite{LL}). 
The paradox
is resolved by finding that a {\em laminar wake} should arise behind 
a moving
body. Within the wake one cannot neglect viscosity even deeply in the 
convection
area when $r \gg r_m$. The width of the laminar wake is growing as 
$\sim \sqrt{r
r_m}$ far from the moving body
\cite{Magnus}. Solving the Navier-Stokes equation by Oseen's method 
one can find out
how the laminar wake and the normal circulation are formed during the 
crossover from
the viscous to the convection subregion. One can find a detailed 
analysis of this
crossover in the presence of the longitudinal and the transverse 
force in Ref.
\cite{Th}. 

The force $\vec F$ is the mutual friction force introduced 
by Hall and Vinen \cite{HV} for the analysis of propagation of the 
second sound in 
rotating superfluid. With help of Eqs. (\ref{Magnus-S}), (\ref{F}), 
and (\ref{f-v}) one obtains 
a linear relation between the mutual friction force and the 
counterflow velocity
$\vec v_s - \vec v_{n\infty}$ bearing in mind that the average normal 
velocity
practically coincides with the velocity $\vec v_{n\infty}$:
\begin{equation}
\vec F={\kappa\rho_s \rho_n \over 2\rho}B (\vec v_s - \vec 
v_{n\infty}) -
{\kappa \rho_s \rho_n \over 2\rho}B'[\hat z \times (\vec v_s - \vec 
v_{n\infty})] ~,  
\label{B-B} \end{equation}
The Hall-Vinen parameters  $B$ and $B'$ are  given by a complex 
relation \cite{RMP}
\begin{equation}
{ 2\rho \over \kappa \rho_s \rho_n }{1\over B-i B'}
=-{1\over i\rho_s \kappa}+{1\over \tilde D + i \tilde D'}=
-{1\over i\rho_s \kappa}+{\ln(r_m/l_{ph})\over 4\pi \eta_n} 
+ {1\over D+i D'}~.
   \label{BB}  \end{equation}

According to this relation a strong viscous drag (large logarithm 
$\ln(r_m/l)$, or
small viscosity $\eta_n$) suppresses the effect of the transverse 
force ($\propto
D'$) on the mutual friction \cite{Th}. But the effect of the 
superfluid Magnus
force and the longitudinal force $\propto D$ is also suppressed in 
this limit. In
fact, this is a limit when the force from quasiparticle scattering 
(transverse, or
longitudinal) is so strong that the normal fluid sticks to the 
vortex, and $\vec
v_L=\vec v_n$. Then the resultant force $\vec F$ depends only on 
viscosity, as in
the classical Stokes problem.

\section{Magnus force and the Berry phase} \label{Berr}

Now let us consider a connection between the transverse force and  
the Berry
phase. We shall use the  hydrodynamic description with the Lagrangian 
obtained 
from Eq. (\ref{Lagr}) after the Madelung transformation:
\begin{equation} 
L={\kappa\rho\over 2\pi }{\partial \phi \over \partial t}-{\kappa^2 
\rho
\over 8\pi }\nabla \phi ^2 -{V\over 2}\rho^2~.
  \label{lagrB}   \end{equation}
The first term with the first time derivative of the phase 
$\phi$ (Wess-Zumino term) is responsible for the Berry phase 
$\Theta =\Delta S_B/\hbar$, which is the variation of the phase of the
quantum-mechanical wave function for an  adiabatic motion of the 
vortex around a
closed loop \cite{B}. Here
\begin{equation} 
\Delta S_B =\int d\vec r\, dt\, {\kappa\rho \over 2\pi }{\partial
\phi \over
\partial t} 
=-\int d\vec r\, dt\, {\kappa\rho \over 2\pi}(\vec v_L \cdot
\vec \nabla_L)\phi~.
    \label{SB}  \end{equation} 
 is the classical action variation around the loop and $\vec 
\nabla_L\phi$ is the
gradient of the phase $\phi[\vec r -\vec r_L(t)]$ with respect to the 
vortex
position vector $\vec r_L(t)$. However, $\vec
\nabla_L\phi= -\vec \nabla \phi$, where $\vec \nabla \phi$ is the 
gradient with respect to
$\vec r$. Then the loop integral $\oint d\vec l$ yields the 
circulation of total current
$\vec j = (\kappa / 2\pi )<\rho \vec \nabla \phi>$ for 
points inside the loop, but vanishes for points outside. As a result, 
the Berry-phase action
is given by \cite{GWT}
\begin{equation} 
\Delta S_B = V{\kappa \over 2\pi}\oint (d\vec
l \cdot \vec j )~.
     \label{jL} \end{equation}
where  $V$ is the volume inside the loop (a product of the loop area 
and the liquid height along a vortex).  Contrary to Eq. (\ref{SB}), 
the loop integral in Eq. (\ref{jL}) is related
with the variation of the position vector $\vec r$, the vortex 
position vector $\vec r_L$
being fixed.

Since the Berry phase is proportional to the current circulation, 
which determines the
transverse force, there is a direct connection between the Berry 
phase and the
amplitude of the transverse force on a vortex, as was shown, e.g., in 
Ref. \cite{AT-N}.Thus in order to find the Berry phase, one should
calculate the current  circulation. If the circulation of the normal
velocity at large distances vanished  (as assumed
in Refs. \cite{AT,AT-N,GWT}),  the current circulation would be
$\oint (d\vec l \cdot \vec j ) =\rho_s \vec \kappa$, and the Berry 
phase (as well
as the effective Magnus  force) would be proportional to $\rho_{s}$. 
However, 
according to Sec. \ref{2D} and and the recent analysis by Thouless {\em
et al.}
\cite{Th}, at very large  distances (in the
convection subregion) the normal circulation $\kappa_n =\oint (d\vec l
\cdot \vec v_n )$ does not vanish and $\oint (d\vec l \cdot \vec j )
=\rho_s \vec \kappa+\rho_n \kappa_n$. Using a proper value of the
asymptotic  normal circulation $\kappa_n$ given by Eq. (\ref{circ}) the
Berry phase yields the same transverse  force as determined from the
momentum balance.

In order to obtain a correct value of the transverse force from the 
Berry phase, one
should choose a loop radius much larger than Oseen's length $r_m$. If 
the loop radius
is chosen in the viscous subregion $l_{ph} < r < r_m$, the 
total-current circulation
is proportional $\rho_s$, but the Berry phase does not yield the 
total transverse
force, since a part of it, namely, the Iordanskii force, is presented 
by the viscous
momentum flux, which cannot be  obtained in the Lagrange formalism. 
And if the loop
radius is chosen in the ballistic region, the total-current 
circulation is not
defined at all and depends on a shape of the loop, since the phonon 
mass current is
not curl-free, as pointed out after Eq. (\ref{mas-vor}).

\section{Discussion and conclusions} \label{C}

The momentum-balance analysis definitely confirms an existence of the
transverse force on a vortex from phonon scattering
(the Iordanskii force). This conclusion agrees with the results of the
recent  analysis of
Thouless {\em et al.} \cite{Th}. 

The Berry phase yields the same value of the transverse force as
the momentum balance, if a proper value of the normal circulation at 
large
distances from the vortex is used for the calculation of the Berry 
phase. However,
the Berry-phase analysis itself cannot provide the normal-circulation 
value, since
the latter is determined by the processes at small distances from the 
vortex,
which are beyond of the Berry-phase analysis. The small-distance 
processes
determine the force between the superfluid and the normal component, 
which is
present in the small-distance boundary condition for the 
Navier-Stokes equation in
the two-fluid-hydrodynamics region. The transverse force on the normal
component  is transmitted
to infinite distances by the constant momentum flux, which requires a 
normal
circulation far from the vortex. This circulation should be used for 
determination
of the Berry phase. 

Ambiguity of the Berry-phase analysis of the transverse force 
originates from   ambiguity of  the transverse force in the Lagrange
formalism,  which was discussed in the end of Sec. V in  Ref.
\cite{PRB7}). Adding a constant $C$ to the total density in the
Wess-Zumino term,  i.e.,  replacing $\rho $ by $\rho +C$, one obtain a
different  amplitudes of the Berry  phase and the transverse force without
any effect  on the field equations for  the condensate wave function (see
also discussion of this constant in Fermi liquids by Volovik \cite{Vol}).
This  arbitrariness has a profound physical meaning.   Derivation of the
Magnus  force in the Lagrange  formalism dealt only with large  distances
much exceeding the vortex  core size. However, the processes inside the
core  affect the total Magnus force in general. Deriving the  Magnus
force from the momentum balance we use the condition that the  total
momentum flux $\int \Pi_{i j}dS_{j}$ through the surface of large radius 
around the  vortex line vanishes. This is true only for a Galilean
invariant  liquid satisfying the momentum conservation law. If the liquid
in the  vortex core interacts with the external world (e.g., with crystal 
impurities in superconductors), the momentum balance condition must 
be  
$\oint  \Pi_{i j}dS_{j}= f_{i}$ where $\vec f$ is the force on the 
vortex core which could also have a transverse component (the 
Kopnin-Kravtsov force 
\cite{KK-S}). One can calculate such a force only from the 
vortex-core 
analysis. The results of this analysis must be used for determination 
of an unknown 
constant in the Wess-Zumino term.

\section*{Acknowledgements}

I thank N. Kopnin, L. Pitaevskii, A. Shelankov, M. Stone, D. 
Thouless, and G.
Volovik for interesting discussions. The work was supported by the 
grant of the Israel
Academy of Sciences and Humanities.

%


\begin{thebibliography}{8.}
\bibitem{Cr-T} But this assumption becomes invalid for the roton 
normal fluid
at higher temperatures, especially close to the critical temperature. 
Then
other methods based on the modified Ginzburg-Pitaevskii theory should 
be used
\cite{RMP,JLT}.
\bibitem{Low} However, scattered
quasiparticles colliding with other quasiparticles immediately after 
scattering
can return back and modify the quasiparticle flux on the vortex. As a 
result of
it, the lower cut-off could be a combination of the  mean-free path
$l_{ph}$ and some effective cross-section of the vortex. This 
logarithmically
weak effect was shortly discussed in the end of Sec. 1 in Ref. 
\cite{S}.
\bibitem{Upp} Note that the choice of
Oseen's length as a upper cut-off of the logarithmic factor is valid 
for an
isolated vortex under a stationary fluid flow. In general the upper 
cut-off
should be the smallest among three scales: Oseen's length, the
frequency-dependent viscous depth, and the intervortex spacing 
\cite{RMP}.
\end{thebibliography}
\end{document}